\documentclass[aps,prb,superscriptaddress,showpacs,floatfix,twocolumn]{revtex4-1}
\usepackage{graphicx,amsmath,amssymb,xspace,epsfig,float,multirow,subfigure,tabularx}

\pdfoutput=1
\input{tcilatex}
\begin{document}

\title{Ginzburg - Landau theory of the triplet superconductivity in 3D Dirac
semi-metal. }
\author{Baruch Rosenstein}
\email{vortexbar@yahoo.com }
\author{B.Ya.Shapiro}
\author{I.Shapiro}
\email{ }
\date{\today }

\begin{abstract}
It was recently shown that conventional phonon-electron interactions induce
triplet pairing states in time-reversal invariant 3D Dirac semi - metals
provided magnetic impurities or exchange interactions are strong enough$^{1}$%
. The order parameter in this case is a vector field. Starting from the
microscopic model of the isotropic Dirac semi-metal, the Ginzburg-Landau
energy for this field is derived using the Gor'kov technique. It was found
that the transversal coherence length $\xi _{T}$ is much smaller than the
longitudinal, $\xi _{L}=4\sqrt{2}\xi _{T},$ despite the isotropy. Several
new features appear when an external field is applied. The Ginzburg - Landau
model is used to investigate magnetic properties of such superconductors.
Using the small deviation method the magnetic penetration depth was found
also to be vastly different for longitudinal and transverse fluctuations $%
\lambda _{T}/\lambda _{L}=4\sqrt{2}$. As a result the superconductor
responds as type I to a transverse perturbation, while with respect to a
longitudinal perturbation it behaves as type II. At large fields the order
parameter orients itself perpendicular to the field direction. The triplet
superconductor persists at arbitrarily high magnetic field (no upper
critical magnetic field) like in some $p$ wave superconductors.
\end{abstract}

\maketitle


\affiliation{Electrophysics Department, National Chiao Tung University, Hsinchu 30050, \textit{Taiwan,
R. O. C}} 
\affiliation{Physics Department, Ariel University, Ariel 40700,
Israel}

\affiliation{Physics Department, Bar-Ilan University, 52900 Ramat-Gan,
Israel}

\affiliation{Physics Department, Bar-Ilan University, 52900 Ramat-Gan,
Israel}


\section{Introduction}

Recently 3D Dirac semi - metals (DSM) like $Na_{3}Bi$ and $Cd_{3}As_{2}$
with electronic states described by Bloch wave functions, obeying the
"pseudo-relativistic" Dirac equation (with the Fermi velocity $v_{F}$
replacing the velocity of light) were observed\cite{Potemski} and attracted
widespread attention. The discovery of the 3D Dirac materials makes it
possible to study their physics including remarkable electronic properties.
This is rich in new phenomena like giant diamagnetism that diverges
logarithmically when the chemical potential approaches the 3D Dirac point, a
linear-in-frequency AC conductivity that has an imaginary part\cite{Wan},
quantum magnetoresistance showing linear field dependence in the bulk\cite%
{Ogata}. Most of the properties of these new materials were measured at
relatively high temperatures. However recent experiments at low temperature
on topological insulators and suspected 3D Dirac semi-metals exhibit
superconductivity. Early attempts to either induce or discover
superconductivity in Dirac matrials were promising. The well known
topological insulator $Bi_{2}Se_{3}$ doped with $Cu$, becomes
superconducting at $T_{c}=3.8K$\cite{Ong}. At present its pairing symmetry
is unknown. Some experimental evidence\cite{phononexp} point to a
conventional phononic pairing mechanism. The spin independent part of the
effective electron - electron interaction due to phonons was studied
theoretically\cite{phonontheory}.\ For a conventional parabolic dispersion
relation, typically independent of spin, the phonon mechanism leads to the $%
s $-wave superconductivity. The layered, non-centrosymmetric heavy element $%
PbTaSe_{2}$ was found to be superconducting \cite{Ong}. Its electronic
properties like specific heat, electrical resistivity, and
magnetic-susceptibility indicate that $PbTaSe_{2}$ is a moderately coupled,
type-II BCS superconductor with large electron-phonon coupling constant of $%
\lambda =0.74$. It was shown theoretically to possess a very asymmetric 3D
Dirac point created by strong spin-orbit coupling. If the 3D is confirmed,
it might indicate that the superconductivity is conventional phonon mediated.

More recently when the $Cu$ doped $Bi_{2}Se_{3}$ was subjected to pressure%
\cite{pressureBiSe}, $T_{c}$ increased to $7K$ at $30GPa$. Quasilinear
temperature dependence of the upper critical field $H_{c2},$ exceeding the
orbital and Pauli limits for the singlet pairing, points to the triplet
superconductivity. The band structure of the superconducting compounds is
apparently not very different from its parent compound $Bi_{2}Se_{3}$, so
that one can keep the two band $\mathbf{k}\cdot \mathbf{p}$ description ($Se$
$p_{z}$ orbitals on the top and bottom layer of the unit cell mixed with its
neighboring $Bi$ $p_{z}$ orbital). Electronic-structure calculations and
experiments on the compounds under pressure\cite{pressureBiSe} reveal a
single bulk three-dimensional Dirac cone like in $Bi$ with large spin-orbit
coupling. Moreover very recently some pnictides were identified as
exhibiting Dirac spectrum\cite{pnictide}. This effort recently culminated in
discovery of superconductivity in $Cd_{3}As_{2}$\cite{3Dsuper}. It is
claimed that the superconductivity is $p$-wave at least on the surface.

The case of the Dirac semi-metals is very special due to the strong spin
dependence of the itinerant electrons' effective Hamiltonian. It was pointed
out\cite{Fu} that in this case the triplet possibility can arise although
the triplet gap is smaller than that of the singlet, the difference
sometimes is not large for spin independent electron - electron
interactions. Very recently the spin dependent part of the phonon induced
electron - electron interaction was considered\cite{DasSarma14} and it was
shown that the singlet is still favored over the triplet pairing. Another
essential spin dependent effective electron-electron interaction is the
Stoner exchange among itinerant electrons leading to ferromagnetism in
transition metals. While in the best 3D Weyl semi-metal candidates it is too
small to form a ferromagnetic state, it might be important to determine the
nature of the superconducting condensate. It turns out that it favors the
triplet pairing\cite{Rosenstein15}. Also a modest concentration of magnetic
impurities makes the triplet ground state stable.

As mentioned above generally the applied magnetic field is an ultimate
technique to probe the superconducting state. In a growing number of
experiments, in addition to magnetotransport, magnetization curves, the
magnetic penetration depth and upper critical magnetic field were measured%
\cite{magnetization}.{\huge \ }It is therefore of importance to construct a
Ginzburg - Landau (GL) description\cite{Ketterson} of these novel materials
to study inhomogeneous order parameter configurations (junctions,
boundaries, etc.) and magnetic properties that typically involve
inhomogeneous configurations (like vortices) not amenable to a microscopic
description.

In the present paper we derive such a GL type theory for triplet
superconductor from the microscopic isotropic DSM model with attractive
local interaction. The order parameter in this case is a vector field and
the GL theory of vector field already considered in literature\cite%
{Knigavko,Machida,Babaev} in connection with putative $p$ - wave
superconductors have several extraordinary features, both quantitative and
qualitative.

The paper is organized as follows. The model of the (phonon mediated or
unconventional) local interactions of 3D Dirac fermion is presented and the
method of its solution (in the Gorkov equations form) including the symmetry
analysis of possible pairing channels and the vectorial nature of the
triplet order parameter is given in Section II. In Section III the Gorkov
formalism, sufficiently general to derive the GL equations, is briefly
presented. The most general form of the GL energy of the triplet
superconductor in magnetic field consistent with the symmetries is given in
IV. The coefficient of the relevant terms are calculated from the
microscopic DSM model in section V. Section VI is devoted to applications of
the GL model. The ground \ state degeneracy, the character of its
excitations and basic magnetic properties are discussed. The vector order
parameter is akin to optical phonons with sharp distinction between
transverse and longitudinal modess. Transverse and longitudinal coherence
lengths and penetration depths are calculated and the upper critical
magnetic field is discussed. Section VI includes generalizations to include
Pauli paramagnetism, discussion of an experimental possibility of
observation of the excitation and\ conclusion.

\section{The local pairing model in the Dirac semi-metal.}

\subsection{Pairing Hamiltonian in the Dirac semi-metal.}

Electrons in the 3D Dirac semimetal are described by field operators $\psi
_{fs}\left( \mathbf{r}\right) $, where $f=L,R$ are the valley index
(pseudospin) for the left/right chirality bands with spin projections taking
the values $s=\uparrow ,\downarrow $ with respect to, for example,\ the $z$
axis. To use the Dirac ("pseudo-relativistic") notations, these are combined
into a four component bi-spinor creation operator, $\psi ^{\dagger }=\left(
\psi _{L\uparrow }^{\dagger },\psi _{L\downarrow }^{\dagger },\psi
_{R\uparrow }^{\dagger },\psi _{R\downarrow }^{\dagger }\right) $, whose
index $\gamma =\left\{ f,s\right\} $ takes four values. The non-interacting
massless Hamiltonian with Fermi velocity $v_{F}$ and chemical potential $\mu 
$ reads\cite{Wang13}%
\begin{eqnarray}
K &=&\int_{\mathbf{r}}\psi ^{+}\left( \mathbf{r}\right) \widehat{K}\psi
\left( \mathbf{r}\right) \text{;\ \ }  \label{kinetic} \\
\text{\ \ \ \ }\widehat{K}_{\gamma \delta } &=&-i\hbar v_{F}\nabla
^{i}\alpha _{\gamma \delta }^{i}-\mu \delta _{\gamma \delta }\text{,}
\end{eqnarray}%
where the three $4\times 4$ matrices, $i=x,y,z$, 
\begin{equation}
\mathbf{\alpha }=\left( 
\begin{array}{cc}
\mathbf{\sigma } & 0 \\ 
0 & -\mathbf{\sigma }%
\end{array}%
\right) \text{,}  \label{alpha}
\end{equation}%
are presented in the block form via Pauli matrices $\mathbf{\sigma }$. They
are related to the Dirac $\mathbf{\gamma }$ matrices (in the chiral
representation, sometimes termed "spinor") by $\ \mathbf{\alpha }=\beta 
\mathbf{\gamma }$ with%
\begin{equation}
\beta =\left( 
\begin{array}{cc}
0 & \mathbf{1} \\ 
\mathbf{1} & 0%
\end{array}%
\right) \text{.}  \label{beta}
\end{equation}%
Here $\mathbf{1}$ is the $2\times 2$ identity matrix.

We consider a special case of 3D rotational symmetry that in particular has
an isotropic Fermi velocity. Moreover we assume time reversal, $\Theta \psi
\left( \mathbf{r}\right) =i\sigma _{y}\psi ^{\ast }\left( \mathbf{r}\right) $%
, and inversion symmetries although the pseudo- Lorentz symmetry will be
explicitly broken by interactions. The spectrum of single particle
excitations is linear. The chemical potential $\mu $ is counted from the
Dirac point.

As usual in certain cases the actual interaction can be approximated by a
model local one:

\begin{equation}
V_{eff}=\mathbf{-}\frac{g}{2}\int_{\mathbf{r}}\psi _{\alpha }^{+}\left( 
\mathbf{r}\right) \psi _{\beta }^{+}\left( \mathbf{r}\right) \psi _{\beta
}\left( \mathbf{r}\right) \psi _{\alpha }\left( \mathbf{r}\right) \text{.}
\label{local}
\end{equation}%
Unlike the free Hamiltonian $K$, Eq.(\ref{kinetic}), this interaction
Hamiltonian does not mix different spin components.

Spin density in Dirac semi-metal has the form 
\begin{equation}
\mathbf{S}\left( \mathbf{r}\right) =\frac{1}{2}\psi ^{+}\left( \mathbf{r}%
\right) \mathbf{\Sigma }\psi \left( \mathbf{r}\right) \text{,}
\label{spindensity}
\end{equation}%
where the matrices 
\begin{eqnarray}
\mathbf{\Sigma } &\mathbf{=-\alpha }&\gamma _{5}=\left( 
\begin{array}{cc}
\mathbf{\sigma } & 0 \\ 
0 & \mathbf{\sigma }%
\end{array}%
\right) ,  \label{sigma_matrices} \\
\gamma _{5} &=&\left( 
\begin{array}{cc}
-\mathbf{1} & 0 \\ 
0 & \mathbf{1}%
\end{array}%
\right) ,
\end{eqnarray}%
are also the rotation generators.

\subsection{The symmetry classification of possible pairing channels.}

Since we consider the local interactions as dominant, the superconducting
condensate (the off-diagonal order parameter) will be local%
\begin{equation}
O=\int_{\mathbf{r}}\psi _{\alpha }^{+}\left( \mathbf{r}\right) M_{\alpha
\beta }\psi _{\beta }^{+}\left( \mathbf{r}\right) ,  \label{O}
\end{equation}%
where the constant matrix $M$ should be a $4\times 4$ antisymmetric matrix.
Due to the rotation symmetry they transform covariantly under infinitesimal
rotations generated by the spin $S^{i}$ operator, Eq.(\ref{spindensity}):

\begin{eqnarray}
&&\int_{\mathbf{r,r}^{\prime }}\left[ \psi _{\alpha }^{+}\left( r\right)
M_{\alpha \beta }\psi _{\beta }^{+}\left( r\right) ,\psi _{\gamma
}^{+}\left( r^{\prime }\right) \mathbf{\Sigma }_{\gamma \delta }^{i}\psi
_{\delta }\left( r^{\prime }\right) \right]  \label{transformation} \\
&=&-\int_{r}\psi _{\gamma }^{+}\left( r\right) \left( \mathbf{\Sigma }%
_{\gamma \delta }^{i}M_{\delta \kappa }+M_{\gamma \delta }\mathbf{\Sigma }%
_{\delta \gamma }^{ti}\right) \psi _{\kappa }^{+}\left( r\right) \text{.}
\end{eqnarray}%
Here and in what follows "$t$" denotes the transpose matrix. The
representations of the rotation group therefore characterize various
possible superconducting phases.

Out of 16 matrices of the four dimensional Clifford algebra six are
antisymmetric and one finds one vector and three scalar multiplets of the
rotation group. The multiplets contain:

(i) a triplet of order parameters: 
\begin{eqnarray}
&&\left\{ M_{x}^{T},M_{y}^{T},M_{z}^{T}\right\}  \label{triplet} \\
&=&\left\{ -\beta \alpha _{z},-i\beta \gamma _{5},\beta \alpha _{x}\right\}
=\left\{ T_{x},T_{y},T_{z}\right\}
\end{eqnarray}%
The algebra is 
\begin{equation}
\mathbf{\Sigma }_{i}T_{j}+T_{j}\mathbf{\Sigma }_{i}^{t}=2i\varepsilon
_{ijk}T_{k}\text{.}  \label{algebra}
\end{equation}%
Note that the three matrices $T_{i}$ are Hermitian.

(ii) three singlets

\begin{equation}
M_{1}^{S}=i\alpha _{y};\text{ \ \ }M_{2}^{S}=i\Sigma _{y};\text{ \ \ }%
M_{3}^{S}=-i\beta \alpha _{y}\gamma _{5}\text{.}  \label{singlets}
\end{equation}%
Which one of the condensates is realized at zero temperature is determined
by the parameters of the Hamiltonian and is addressed next within the
Gaussian approximation. As was shown in our previous work\cite{Rosenstein15}
either exchange interactions or magnetic impurities make the triplet state a
leading superconducting channel in these materials. Therefore we will
consider in the next section only the vector channel.

\section{Gorkov equations and the triplet pairing}

\subsection{Gorkov equations for Green's functions in matrix form}

Using the standard BCS formalism, the Matsubara Green's functions ($\tau $
is the Matsubara time) 
\begin{eqnarray}
G_{\alpha \beta }\left( \mathbf{r},\tau ;\mathbf{r}^{\prime },\tau ^{\prime
}\right) &=&-\left\langle T_{\tau }\psi _{\alpha }\left( \mathbf{r},\tau
\right) \psi _{\beta }^{\dagger }\left( \mathbf{r}^{\prime },\tau ^{\prime
}\right) \right\rangle \text{;}  \label{GFdef} \\
F_{\alpha \beta }\left( \mathbf{r},\tau ;\mathbf{r}^{\prime },\tau ^{\prime
}\right) &=&\left\langle T_{\tau }\psi _{\alpha }\left( \mathbf{r},\tau
\right) \psi _{\beta }\left( \mathbf{r}^{\prime },\tau ^{\prime }\right)
\right\rangle \text{;}  \notag \\
F_{\alpha \beta }^{+}\left( \mathbf{r},\tau ;\mathbf{r}^{\prime },\tau
^{\prime }\right) &=&\left\langle T_{\tau }\psi _{\alpha }^{\dagger }\left( 
\mathbf{r},\tau \right) \psi _{\beta }^{\dagger }\left( \mathbf{r}^{\prime
},\tau ^{\prime }\right) \right\rangle \text{,}  \notag
\end{eqnarray}%
obey the Gor'kov equations\cite{AGD}:%
\begin{gather}
-\frac{\partial G_{\gamma \kappa }\left( \mathbf{r},\tau ;\mathbf{r}^{\prime
},\tau ^{\prime }\right) }{\partial \tau }-\int_{\mathbf{r}^{\prime \prime
}}\left\langle \mathbf{r}\left\vert \widehat{K}_{\gamma \beta }\right\vert 
\mathbf{r}^{\prime \prime }\right\rangle G_{\beta \kappa }\left( \mathbf{r}%
^{\prime \prime },\tau ;\mathbf{r}^{\prime },\tau ^{\prime }\right)
\label{Gorkov} \\
-gF_{\beta \gamma }\left( \mathbf{r},\tau ;\mathbf{r},\tau \right) F_{\beta
\kappa }^{+}\left( \mathbf{r},\tau ,\mathbf{r}^{\prime },\tau ^{\prime
}\right) =\delta ^{\gamma \kappa }\delta \left( \mathbf{r-r}^{\prime
}\right) \delta \left( \tau -\tau ^{\prime }\right) ;  \notag \\
\frac{\partial F_{\gamma \kappa }^{+}\left( \mathbf{r},\tau ;\mathbf{r}%
^{\prime },\tau ^{\prime }\right) }{\partial \tau }-\int_{\mathbf{r}^{\prime
\prime }}\left\langle \mathbf{r}\left\vert \widehat{K}_{\gamma \beta
}^{t}\right\vert \mathbf{r}^{\prime \prime }\right\rangle F_{\beta \kappa
}^{+}\left( \mathbf{r}^{\prime \prime },\tau ;\mathbf{r}^{\prime },\tau
^{\prime }\right)  \notag \\
-gF_{\gamma \beta }^{+}\left( \mathbf{r},\tau ;\mathbf{r},\tau \right)
G_{\beta \kappa }\left( \mathbf{r},\tau ,\mathbf{r}^{\prime },\tau ^{\prime
}\right) =0\text{.}  \notag
\end{gather}%
These equations are conveniently presented in matrix form (superscript $t$
denotes transposed and $I$ - the identity matrix): 
\begin{eqnarray}
\int_{X^{\prime \prime }}D^{-1}\left( X,X^{\prime \prime }\right) G\left(
X^{\prime \prime },X^{\prime }\right) -\Delta \left( X\right) F^{+}\left(
X,X^{\prime }\right) &=&I\delta \left( X-X^{\prime }\right) \text{;}
\label{matrixeq} \\
\int_{X^{\prime \prime }}D^{t-1}\left( X,X^{\prime \prime }\right)
F^{+}\left( X^{\prime \prime },X^{\prime }\right) +\Delta ^{t\ast }\left(
X\right) G\left( X,X^{\prime }\right) &=&0\text{.}  \notag
\end{eqnarray}%
Here $X=\left( \mathbf{r},\tau \right) $, $\Delta _{\alpha \beta }\left(
X\right) =gF_{\beta \alpha }\left( X,X\right) $ and 
\begin{equation}
D_{\alpha \beta }^{-1}\left( X,X^{\prime }\right) =-\delta _{\alpha \beta }%
\frac{\partial }{\partial \tau }\delta \left( X-X^{\prime }\right) -\delta
\left( \tau -\tau ^{\prime }\right) \left\langle \mathbf{r}\left\vert 
\widehat{K}_{\alpha \beta }\right\vert \mathbf{r}^{\prime }\right\rangle 
\text{.}  \label{invD}
\end{equation}

In the homogeneous case the Gor'kov equations for Fourier components of the
Green's functions simplify considerably:

\begin{eqnarray}
D^{-1}\left( \omega ,\mathbf{p}\right) G\left( \omega ,\mathbf{p}\right)
-\Delta F^{+}\left( \omega ,\mathbf{p}\right) &=&I\text{;}
\label{Gorkov_uniform} \\
\widetilde{D}^{-1}\left( \omega ,\mathbf{p}\right) F^{+}\left( \omega ,%
\mathbf{p}\right) +\Delta ^{t\ast }G\left( \omega ,\mathbf{p}\right) &=&0%
\text{.}  \notag
\end{eqnarray}%
The matrix gap function can be chosen as 
\begin{equation}
\Delta _{\beta \gamma }=gF_{\gamma \beta }\left( 0\right) =\Delta
_{z}M_{\gamma \beta }\text{,}  \label{delta}
\end{equation}%
with real constant $\Delta _{z}$. Here $D^{-1}\left( \omega ,\mathbf{p}%
\right) =i\omega +\mu -\mathbf{\alpha \cdot p}$, is the noninteracting
inverse Dirac Green's function for the Hamiltonian Eq.(\ref{kinetic}) and $%
\widetilde{D}^{-1}\left( \omega ,\mathbf{p}\right) =i\omega -\mu -\mathbf{%
\alpha }^{t}\mathbf{\cdot p}$, where $\omega _{n}=\pi T\left( 2n+1\right) $
is the fermionic Matsubara frequency.

Solving these equations one obtains (in matrix form) 
\begin{eqnarray}
G^{-1} &=&D^{-1}+\Delta \widetilde{D}\Delta ^{t\ast }\text{;}
\label{solution} \\
F^{+} &=&-\widetilde{D}\Delta ^{t\ast }G\text{,}  \notag
\end{eqnarray}%
with the gap function to be found from the consistency condition 
\begin{equation}
\Delta ^{t\ast }=-g\sum\limits_{\omega p}\widetilde{D}\Delta ^{t\ast }G\text{%
.}  \label{gap eq}
\end{equation}%
Now we find solutions of this equation for each of the possible
superconducting phases.

\subsection{Homogeneous triplet solution of the gap equation.}

In this phase rotational symmetry is spontaneously broken simultaneously
with the electric charge $U\left( 1\right) $ (global gauge invariance)
symmetry. Assuming $z$ direction of the $p$ - wave condensate the order
parameter matrix takes a form: 
\begin{equation}
\Delta =\Delta _{z}T_{z}=\Delta _{z}\beta \alpha _{x},  \label{deltaT1}
\end{equation}%
where $\Delta _{z}$ is a constant.{\huge \ }The energy scale will be set by
the Debye cutoff $T_{D}$ of the electron - phonon interactions, see below.

The spectrum of elementary excitations at zero temperature was discussed in
ref. \cite{Rosenstein15}. There is a saddle point with energy gap $2\Delta
_{z}$ on the circle $p_{\perp }^{2}\equiv p_{x}^{2}+p_{y}^{2}=\mu
^{2}/v_{F}^{2},$ $p_{z}=0$. The gap $\Delta _{z}$ as a function of the
dimensionless phonon-electron coupling $\lambda =gN$, where $N$ being the
density of states (all spins and valleys included), increases upon reduction
in $\mu $. At large $\mu >>T_{D}$, as in BCS, the gap becomes independent of 
$\mu $ and one has the relation%
\begin{equation}
\frac{1}{g}=\frac{N}{12}\sinh ^{-1}\frac{T_{D}}{\Delta _{z}};N=\frac{2\mu
^{2}}{\pi ^{2}v_{F}^{3}\hbar ^{3}}\text{,}  \label{gapBCS_T}
\end{equation}%
leading to an exponential gap dependence on $\lambda $ when it is small: $%
\Delta _{z}=T_{D}/\sinh \left( 12/\lambda \right) \simeq
2T_{D}e^{-12/\lambda }$.

The critical temperature is obtained from Eq.(\ref{gap eq}) with
disctretized $\omega $ by substituting $\Delta _{z}=0$. To utilize the
orthonormality of $T_{i}$, Tr$\left( T_{i}T_{j}^{\ast }\right) =4\delta
_{ij} $, one multiplies the gap equation by the matrix $T_{z}/g$ and takes
the trace: 
\begin{equation}
\frac{1}{g}\text{Tr}\left( T_{z}T_{z}^{\ast }\right) =\frac{4}{g}=T_{c}%
\mathcal{B}_{zz}\text{.}  \label{Tc}
\end{equation}%
The bubble integral is 
\begin{eqnarray}
\mathcal{B}_{ij} &=&\sum\limits_{\omega \mathbf{p}}\text{Tr}\left( T_{i}%
\widetilde{D}T_{j}^{\ast }D\right) =4\delta _{ij}T_{c}\times  \label{Bij} \\
&&\times \sum\limits_{n\mathbf{p}}\frac{v_{F}^{2}\left( p_{\perp
}^{2}-p_{z}^{2}\right) +\mu ^{2}+\omega _{n}^{2}}{\omega _{n}^{4}+\left(
v_{F}^{2}p^{2}-\mu ^{2}\right) ^{2}+2\omega _{n}^{2}\left(
v_{F}^{2}p^{2}+\mu ^{2}\right) }\text{.}
\end{eqnarray}%
Performing first the sum over Matsubara frequencies and then integrate over $%
q$ one obtains, similarly to the singlet BCS, (see Appendix A for details):

\begin{equation}
T_{c}=\frac{2\gamma _{E}}{\pi }T_{D}e^{-12/\lambda }\text{,}  \label{Tcres}
\end{equation}%
where $\log \gamma _{E}=0.577$ is the Euler constant.

\section{A general GL description of a triplet superconductor in a magnetic
field.}

In this section the effective description of the superconducting condensate
in terms of the varying (on the mesoscopic scale) order complex parameter
vector field $\Delta _{i}\left( \mathbf{r}\right) $ is presented.

\subsection{The GL description for a vector order parameter}

The static phenomenological description is determined by the GL free energy
functional $F\left[ \mathbf{\Delta }\left( \mathbf{r}\right) ,\mathbf{A}%
\left( \mathbf{r}\right) \right] $ expanded to second order in gradients and
fourth order in $\Delta $. In a magnetic field $\mathbf{B}$, as usual, space
derivatives of the microscopic Hamiltonian become covariant derivatives $%
\mathbf{\nabla \rightarrow }\mathcal{D}\mathbf{=\nabla +}i\frac{e^{\ast }}{%
\hbar c}\mathbf{A}$, $e^{\ast }=2e$ due to gauge invariance under $\Delta
_{i}\rightarrow e^{i\chi \left( \mathbf{r}\right) }\Delta
_{i},A_{i}\rightarrow A_{i}-\frac{\hbar c}{e^{\ast }}\nabla \chi $. Naively
the only modification of the GL energy is in the gradient term, Eq.(\ref%
{gradterms}); the most general gradient term consistent with rotation
symmetry and the $U\left( 1\right) $ gauge symmetry is 
\begin{equation}
F_{\func{grad}}=N\int_{\mathbf{r}}\left\{ 
\begin{array}{c}
u_{T}\left\{ \left( \mathcal{D}_{j}\Delta _{i}\right) ^{\ast }\left( 
\mathcal{D}_{j}\Delta _{i}\right) -\left( \mathcal{D}_{i}\Delta _{j}\right)
^{\ast }\left( \mathcal{D}_{j}\Delta _{i}\right) \right\} \\ 
+u_{L}\left( \mathcal{D}_{i}\Delta _{j}\right) ^{\ast }\left( \mathcal{D}%
_{j}\Delta _{i}\right)%
\end{array}%
\right\} \text{.}  \label{gradterms}
\end{equation}

The factor $N,$ the density of states, is customarily introduced into energy 
\cite{Rosenstein15}. It was noted in \cite{Machida}, that, unlike in the
usual scalar order parameter case, the longitudinal and transverse
coefficients are in general different, leading to two distinct coherence
lengths, see Section V. Possibilities for the local terms are\cite{Knigavko}%
\begin{equation}
F_{loc}=N\left( \mu \right) \int_{\mathbf{r}}\left\{ \alpha \left(
T-T_{c}\right) \Delta _{i}^{\ast }\Delta _{i}+\frac{\beta _{1}}{2}\left(
\Delta _{i}^{\ast }\Delta _{i}\right) ^{2}+\frac{\beta _{2}}{2}\left\vert
\Delta _{i}\Delta _{i}\right\vert ^{2}\right\} \text{.}  \label{potterms}
\end{equation}

The magnetic part, $F_{mag}=B^{2}/8\pi $, completes the free energy.

\subsection{The set of GL equations for triplet order parameter}

The set of the GL equations corresponding to this energy are obtained by
variation with respect to $\Psi _{j}^{\ast }$ and $A_{i}$. The first is:

\begin{equation}
-\left\{ u_{T}\left( \delta _{ij}\mathcal{D}^{2}-\frac{1}{2}\left\{ \mathcal{%
D}_{i},\mathcal{D}_{j}\right\} \right) +\frac{1}{2}u_{L}\left\{ \mathcal{D}%
_{i},\mathcal{D}_{j}\right\} \right\} \Delta _{j}+  \label{GLEZ}
\end{equation}%
\begin{equation*}
+\alpha \left( T-T_{c}\right) \Delta _{i}+\beta _{1}\Delta _{i}\Delta
_{j}^{\ast }\Delta _{j}+\beta _{2}\Delta _{i}^{\ast }\Delta _{j}\Delta _{j}=0%
\text{.}
\end{equation*}%
The anticommutator appears due complex conjugate terms in Eq.(\ref{gradterms}%
)\cite{Maki}. The Maxwell equation for the supercurrent density is:

\begin{equation}
\ J_{i}=\frac{ie^{\ast }}{\hbar \ }N\left( \mu \right) \left\{ u_{T}\Delta
_{j}^{\ast }\mathcal{D}_{i}\Delta _{j}+u\Delta _{j}^{\ast }\mathcal{D}%
_{j}\Delta _{i}\right\} +cc\text{,}  \label{current}
\end{equation}%
where $u=u_{L}-u_{T}$.

Having established coefficients $u_{T,L}$,$\beta _{1,2}$, $\mu _{Z}$ and $%
\alpha $, our aim in the next Section is to deduce them from the microscopic
Dirac semi-metal model.

\section{GL coefficients from the Gor'kov equations}

For the calculation of coefficients of the local part, the homogeneous
Gor'kov equation, Eq.(\ref{gap eq}) suffices, while for calculation of the
gradient terms a general linearized equation, Eq.(\ref{matrixeq}) is
necessary.{\huge \ }Magnetic field effects are required only for the
calculation of the Zeeman term coefficients and normalization of the order
parameter via supercurrent density.

\subsection{Local (potential) terms in Gor'kov}

Iterating once the equation Eq.(\ref{gap eq}) with help of Eq.(\ref{solution}%
) one obtains the local terms to third order in the gap function:

\begin{equation}
\frac{1}{g}\Delta ^{\ast t}+\sum\limits_{\omega \mathbf{p}}\left\{ 
\begin{array}{c}
\widetilde{D}\left( \omega ,\mathbf{p}\right) \Delta ^{\ast t}D\left( \omega
,\mathbf{p}\right) - \\ 
-\widetilde{D}\left( \omega ,\mathbf{p}\right) \Delta ^{\ast t}D\left(
\omega ,\mathbf{p}\right) \Delta \widetilde{D}\left( \omega ,\mathbf{p}%
\right) \Delta ^{\ast t}D\left( \omega ,\mathbf{p}\right)%
\end{array}%
\right\} \text{.}  \label{expanded}
\end{equation}%
Using $\Delta ^{t\ast }=\Delta _{i}^{\ast }T_{i}$, multiplying by $T_{i}^{t}$
and taking the trace, one gets the linear local terms%
\begin{equation}
N\alpha \left( T-T_{c}\right) \Delta _{i}^{\ast }=\frac{4}{g}\Delta
_{i}^{\ast }\text{,}  \label{linear1}
\end{equation}%
where the bubble integral was given in Eq.(\ref{Bij}). Expressing $g$ via $%
T_{c}$, see Eq.(\ref{Tcres}) allows to write the coefficient $a$ of $\Delta
_{i}^{\ast }$ in the Gorkov equation Eq.(\ref{gap eq}) is%
\begin{equation}
\alpha \left( T-T_{c}\right) =\frac{8\mu ^{2}}{3\pi ^{2}v_{F}^{3}\hbar ^{3}N}%
\log \frac{T}{T_{c}}\approx \frac{4}{3}\frac{T-T_{c}}{T_{c}}\text{,}
\label{a(T)}
\end{equation}%
where $N$ is the density of states.

The cubic terms in Eq.(\ref{expanded}), multiplied again by $T_{i}^{t}$ and
"traced" take the form 
\begin{eqnarray}
&&N\left( \beta _{1}\Delta _{j}^{\ast }\Delta _{j}\Delta _{i}^{\ast }+\beta
_{2}\Delta _{j}^{\ast }\Delta _{j}^{\ast }\Delta _{i}\right)  \label{terms}
\\
&=&-\Delta _{j}^{\ast }\Delta _{k}\Delta _{l}^{\ast }\dsum\limits_{\omega 
\mathbf{p}}\text{Tr}\left\{ T_{i}^{t}\widetilde{D}T_{j}^{t\ast }DT_{k}%
\widetilde{D}T_{l}^{t\ast }D\right\} \text{.}
\end{eqnarray}%
The calculation is given in Appendix A and results in: 
\begin{equation}
\beta _{1}=\frac{7\zeta \left( 3\right) }{20\pi ^{2}}\frac{1}{T_{c}^{2}};%
\text{ \ \ }\beta _{2}=-\frac{1}{3}\beta _{1}\text{.}  \label{betas}
\end{equation}%
The Riemann zeta function is $\zeta \left( 3\right) =1.\,2\ $.

\subsection{Linear gradient terms}

To calculate the gradient terms, one first linearizes the Gor'kov equations,
Eq.(\ref{matrixeq})%
\begin{eqnarray}
&&\int_{X^{\prime \prime }}D^{-1}\left( X,X^{\prime \prime }\right) G\left(
X^{\prime \prime },X^{\prime }\right)  \label{Gorkovlin} \\
&=&I\delta \left( X-X^{\prime }\right) \rightarrow G=D^{-1}\text{;} \\
&&F^{+}\left( X,X^{\prime }\right)  \notag \\
&=&-\int_{X^{\prime \prime }}D^{t}\left( X-X^{\prime \prime }\right) \Delta
^{\ast t}\left( X^{\prime \prime }\right) D\left( X^{\prime \prime
}-X^{\prime }\right) \text{.}
\end{eqnarray}%
In particular,%
\begin{eqnarray}
\frac{1}{g}\Delta ^{\ast t}\left( X\right) &=&F^{+}\left( X,X\right)
\label{DeltaX} \\
&=&-\int_{X^{\prime }}D^{t}\left( X-X^{\prime }\right) \Delta ^{\ast
t}\left( X^{\prime }\right) D\left( X^{\prime }-X\right) \text{.}
\end{eqnarray}%
The anomalous Green's functions are no longer space translation invariant,
so that the following Fourier transform is required: The (time independent)
order parameter is also represented via Fourier components $\Delta ^{\ast
}\left( X\right) =\sum_{\mathbf{P}}e^{-i\mathbf{P}\cdot \mathbf{r}}\Delta
^{\ast }\left( \mathbf{P}\right) $. The linear part Gor'kov equation (this
time including nonlocal parts) {\huge as} the only "external" momentum $%
\mathbf{P}$ reads%
\begin{equation}
\frac{1}{g}\Delta ^{\ast t}\left( \mathbf{P}\right) +\sum_{\omega \mathbf{p}}%
\widetilde{D}\left( \omega ,\mathbf{p}\right) \Delta ^{t\ast }\left( \mathbf{%
P}\right) D\left( \omega ,\mathbf{p-P}\right) \text{.}  \label{DeltaP}
\end{equation}

To find the coefficients of the gradient terms, one should consider
contributions quadratic in $P$ from the expansion of both $\Delta ^{t\ast
}\left( \mathbf{P}\right) $ and $D\left( \omega ,\mathbf{p-P}\right) $. In
view of the gap equation Eq.(\ref{Tc},\ref{Bij}), the expansions of $\Delta
^{t\ast }\left( \mathbf{P}\right) $ cancel each other up to small
corrections of order $T-T_{c}$. So that multiplying by $\,T_{i}^{t}$ and
taking the trace

\begin{equation}
\frac{1}{2}P_{k}P_{l}\sum_{\omega \mathbf{p}}\text{Tr}\left\{ T_{i}^{t}%
\widetilde{D}\left( \omega ,\mathbf{p}\right) T_{j}^{t\ast }D_{kl}^{\prime
\prime }\left( \omega ,\mathbf{p}\right) \right\} \Delta _{j}^{\ast }\text{,}
\label{nonlocal _ terms}
\end{equation}%
where $D_{kl}^{\prime \prime }\left( \omega ,\mathbf{p}\right) \equiv \frac{%
\partial ^{2}D\left( \omega ,\mathbf{p}\right) }{\partial p_{k}\partial p_{l}%
}$. Comparing this with the gradient terms in the GL equation, Eq.(\ref{GLEZ}%
), see Appendix B for details, one deduces

\begin{equation}
u_{T}=\frac{28\zeta \left( 3\right) }{15\pi ^{2}}\frac{v_{F}^{2}\hbar ^{2}}{%
T_{c}^{2}};\text{ \ }u_{L}=\frac{1}{32}u_{T}\text{.}  \label{u_res}
\end{equation}%
Note the very small longitudinal coefficient, $u_{L}<<u_{T}$. As we shall
see in the following section it has profound phenomenological consequences.

\section{Basic properties of the triplet superconductor}

\subsection{Ground state structure and degeneracy}

A ground state is characterized by three independent parameters
corresponding to three Goldstone bosons. The GL energy is invariant under
both the vector $SO(3)$ rotations, $\Delta _{i}\rightarrow R_{ij}\Delta _{j}$%
, and the superconducting phase $U(1)$, $\Delta _{i}\rightarrow e^{i\chi
}\Delta _{i}$. In the superconducting state characterized by the vector
order parameter $\mathbf{\Delta }$ ($\left\vert \mathbf{\Delta }\right\vert
=\Delta $, energy gap) the $U\left( 1\right) $ is broken: $U\left( 1\right)
\rightarrow 1$, while the $SO(3)$ is only partially broken down to its $%
SO(2) $. There are therefore three Goldstone modes. Here we explicitly
parametrize these degrees of freedom by phases following ref.\cite{Knigavko}%
. Generally a complex vector field can be written as

\begin{equation}
\mathbf{\Delta }=\Delta \left( \mathbf{n}\cos \phi \mathbf{+}i\mathbf{m}\sin
\phi \right) ,  \label{parametrization}
\end{equation}%
where $\mathbf{n}$ and $\mathbf{m}$ are arbitrary unit vectors and $0<\phi
<\pi /2$. Using this parametrization the homogeneous part of the free-energy
density, Eq.(\ref{potterms}), takes the form

\begin{equation}
f_{loc}=N\left\{ 
\begin{array}{c}
\alpha \left( T-T_{c}\right) \Delta ^{2} \\ 
+\frac{1}{2}\left( \beta _{1}+\beta _{2}\left( \cos ^{2}\left( 2\phi \right)
+\left( \mathbf{n\cdot m}\right) ^{2}\sin ^{2}\left( 2\phi \right) \right)
\right) \Delta ^{4}%
\end{array}%
\right\} \text{.}  \label{potential}
\end{equation}

This form allows us to make several interesting observations. The crucial
sign is that of $\beta _{2}$. In previous studies\cite{Knigavko,Bel} only $%
\beta _{2}>0$ (so called phase A) was considered. In our case however $\beta
_{2}<0$ and different ground state configurations should be considered. In
phase B the minimization gives, $\mathbf{n=\pm m}$. Note two different
solutions. So that the "vacuum manifold" is 
\begin{equation}
\mathbf{\Delta }=\Delta _{0}\mathbf{n}e^{i\chi }\text{.}  \label{cacuum}
\end{equation}%
Here the range of $\chi $ was enlarged, $-\pi /2<\chi <\pi /2$, to
incorporate $\mathbf{n=\pm m}$. One can use the spherical representation of
the unit vector (that will be termed "director"): $\mathbf{n}=\left( \sin
\theta _{n}\cos \varphi _{n},\sin \theta _{n}\sin \varphi _{n},\cos \theta
_{n}\right) $. The ground state energy density therefore is achieved at 
\begin{equation}
\Delta _{0}^{2}=\frac{\alpha \left( T_{c}-T\right) }{\beta _{1}+\beta _{2}}%
\text{.}  \label{delta0}
\end{equation}%
Mathematically the vacuum manifold in phase B is isomorphic to $S_{2}\otimes
S_{1}/Z_{2}$. This determines the thermodynamics of the superconductor very
much in analogy with the scalar superconductor with $\beta =\beta _{1}+\beta
_{2}$. Magnetic properties are however markedly different.

\subsection{Small fluctuations analysis: two vastly different penetration
depths and coherence lengths.}

Here the response of the superconductor in phase B to an external
perturbation, like boundary or magnetic field, is considered. The basic
excitation modes are uncovered by the linear stability analysis very similar
to the so-called Anderson - Higgs mechanism in field theory applied to
(scalar order parameter) superconductivity a long time ago\cite{Weinberg}.
Two basic scales, the coherence length (scale of variations of the order
parameter) and magnetic penetration depth (scale of variations of the
magnetic field),{\huge \ }are obtained from the expansion of the GL energy
to second order in fluctuations around superconducting ground state at zero
field. The fluctuations are parametrized by $\theta _{n},\varphi _{n},\theta
_{m},\varphi _{m},\phi ,A_{i}$, $\Delta =\Delta _{0}\left( 1+\varepsilon
\right) $. The order parameter, Eq.(\ref{parametrization}), to the second
order is 
\begin{eqnarray}
\mathbf{\Delta /}\Delta _{0} &\approx &\left( 0,0,1\right) +\left( \theta
_{n},0,\varepsilon +i\phi \right)  \label{orderparameterexp} \\
&&+\left( \varepsilon \theta _{n}+i\theta _{m}\phi ,\theta _{n}\varphi
_{n}-\theta _{n}^{2}/2+i\varepsilon \phi -\phi ^{2}/2\right) \text{.}
\end{eqnarray}%
For a perturbation with wave vector $\mathbf{k}$ the energy to quadratic
order in fluctuations reads:%
\begin{equation}
\frac{\delta F}{N}=\sum\nolimits_{\mathbf{k}}\left\{ 
\begin{array}{c}
\frac{\alpha \left( T_{c}-T\right) }{\beta _{1}+\beta _{2}}\left\{ 2\alpha
\left( T_{c}-T\right) \varepsilon ^{\ast }\varepsilon +u_{T}W_{T}+uW\right\}
\\ 
+\frac{\hbar ^{2}c^{2}}{8\pi e^{\ast 2}N}A_{i}^{\ast }\left( k^{2}\delta
_{ij}-k_{i}k_{j}\right) A_{j}%
\end{array}%
\right\} \text{,}  \label{harm}
\end{equation}%
where%
\begin{eqnarray}
W_{T}^{{}} &=&k^{2}\left( \theta _{n}^{\ast }\theta _{n}+\varepsilon ^{\ast
}\varepsilon +\phi ^{\ast }\phi +i\left( \varepsilon ^{\ast }\phi -\phi
^{\ast }\varepsilon \right) \right)  \label{W} \\
&&+\left( \varepsilon ^{\ast }-i\phi ^{\ast }\right)
k_{j}A_{j}+k_{j}A_{j}^{\ast }\left( \varepsilon +i\phi \right) +A_{j}^{\ast
}A_{j}\text{;} \\
W &=&k_{1}^{2}\theta _{n}^{\ast }\theta _{n}+k_{1}k_{3}\theta _{n}^{\ast
}\left( \varepsilon +i\phi \right) +k_{1}k_{3}\left( \varepsilon ^{\ast
}-i\phi ^{\ast }\right) \theta _{n}  \notag \\
&&+k_{3}^{2}\left( \varepsilon ^{\ast }\varepsilon +\phi ^{\ast }\phi
+i\left( \varepsilon ^{\ast }\phi -\phi ^{\ast }\varepsilon \right) \right)
\\
&&+k_{3}\left( 
\begin{array}{c}
\theta _{n}^{\ast }A_{1}+\left( \varepsilon ^{\ast }-i\phi ^{\ast }\right)
A_{3}+A_{1}^{\ast }\theta _{n}^{{}} \\ 
+A_{3}^{\ast }\left( \varepsilon +i\phi \right)%
\end{array}%
\right) +A_{3}^{\ast }A_{3}\text{.}  \notag
\end{eqnarray}%
Diagonalization of this quadratic form reveals three Goldstone modes (that
do not affect the characteristic lengths) and the "massive" fields, $%
\varepsilon $ and $\mathbf{A}$\ (the only one contributing for $\mathbf{k}=0$%
) with different longitudinal and transversal characteristic lengths:%
\begin{eqnarray}
\xi _{T}^{2} &=&\frac{u_{T}}{2\alpha \left( T_{c}-T\right) }\text{; \ \ \ \
\ \ \ \ \ }\xi _{L}^{2}=\text{ }\frac{u_{L}}{u_{T}}\xi _{T}^{2}\text{;}
\label{xipar} \\
\lambda _{T}^{2} &=&\frac{\hbar ^{2}c^{2}\left( \beta _{1}+\beta _{2}\right) 
}{8\pi e^{\ast 2}u_{T}\alpha \left( T_{c}-T\right) N}\text{; \ \ }\lambda
_{L}^{2}=\text{ }\frac{u_{T}}{u_{L}}\lambda _{T}^{2}\text{.}  \notag
\end{eqnarray}%
Here $\xi _{T}$ is the coherence length along the directions perpendicular
to the vector $\mathbf{n}$, while the one parallel to $\mathbf{n}$ is $\xi
_{L}$.

Similarly for magnetic penetration depths $\lambda _{T,L}$. Our calculation
in the previous Section for the Dirac semi-metal, see Eq.(\ref{u_res}),
demonstrate that both are quite different since $u_{L}/u_{T}=1/32<<1$. This
is obviously of great importance for large magnetic field properties of such
superconductors. The phenomenological consequences for vortex state are
briefly discussed in Section VI.

\subsection{Strong magnetic fields: is there an upper critical field $H_{c2}$%
?}

In strong homogeneous magnetic field $H$ (assumed to be directed along the $%
z $ axis) superconductivity typically (but not always, see an example of the 
$p $-wave superconductor that develops flux phases \cite{pwave}) disappears
at certain critical value $H_{c2}$. This bifurcation point is determined
within the GL framework by the lowest eigenvalue of the linearized GL
equations. This is an exact requirement of stability of the normal phase\cite%
{Ketterson,Alama}. The linearized GL equation Eq.(\ref{GLEZ}) reads:%
\begin{equation}
\left[ \left( a-u_{T}\mathcal{D}^{2}\right) \delta _{ij}-\frac{u}{2}\left\{ 
\mathcal{D}_{i},\mathcal{D}_{j}\right\} \right] \Delta _{j}=0\text{,}
\label{linearized GL}
\end{equation}%
where coefficients are in Eq.(\ref{u_res}), and $u=u_{L}-u_{T}$. We use the
Landau gauge, $A_{x}=H_{c2}y;$ $A_{y}=A_{z}=0$. Assuming translation
symmetry along the field direction, $\partial _{z}\Delta _{i}=0$, the
operators of the eigenvalue problem depend on $x$ and $y$ only.

Since we have three components of the order parameter, there are three
eigenvalues. It is easily seen from Eq.(\ref{linearized GL}) that the $z$-
component of the order parameter $\Delta _{z}$ parallel to the external
field direction is independent of the other two, $\Delta _{x},\Delta _{y}$,
leading to the ordinary Abrikosov value:%
\begin{equation}
-u_{T}\mathcal{D}^{2}\Delta _{z}=-a\Delta _{z}\rightarrow H_{c2}^{\parallel
}=\frac{\Phi _{0}}{2\pi \xi _{T}^{2}}\text{,}  \label{transversalHc2}
\end{equation}%
\ where $\xi _{T}$ is defined in Eq.(\ref{xipar}). To avoid confusion with
customary notations for layered materials (like high $T_{c}$ cuprates), the
material that is modelled here is isotropic and "parallel", "perpendicular"
and refer to the relative orientation of the magnetic field to the vector
order parameter rather than to a layer. The orientation of the order
parameter in isotropic material considered here, due to degeneracy of the
ground state, is determined by the external magnetic field as we exemplify
next.

The two remaining eigenvalues\ involving only the order parameter components 
$\Delta _{x}$ and $\Delta _{y}$ perpendicular to the field are obtained from
diagonalizing the "Hamiltonian":

\begin{eqnarray}
\mathcal{H}\left( 
\begin{array}{c}
\Delta _{x} \\ 
\Delta _{y}%
\end{array}%
\right) &=&-a\left( 
\begin{array}{c}
\Delta _{x} \\ 
\Delta _{y}%
\end{array}%
\right) ;  \label{2DVar} \\
\mathcal{H} &\mathcal{=}&\mathcal{-}\left( 
\begin{array}{cc}
u_{T}\mathcal{D}_{y}^{2}+u_{L}\mathcal{D}_{x}^{2} & \frac{u}{2}\left\{ 
\mathcal{D}_{x},\mathcal{D}_{y}\right\} \\ 
\frac{u}{2}\left\{ \mathcal{D}_{x},\mathcal{D}_{y}\right\} & u_{T}\mathcal{D}%
_{x}^{2}+u_{L}\mathcal{D}_{y}^{2}%
\end{array}%
\right) \text{.}  \notag
\end{eqnarray}%
This nontrivial eigenvalue problem fortunately can be solved exactly, see
Appendix C. The lowest eigenstate being a superposition of just two lowest
even Landau levels, $\left\vert 0\right\rangle $ and $\left\vert
2\right\rangle $ are: $\Delta _{x}=\alpha \left\vert 0\right\rangle +\beta
\left\vert 2\right\rangle $, $\Delta _{y}=\gamma \left\vert 0\right\rangle
+i\beta \left\vert 2\right\rangle $. The lowest of these eigenvalues is%
\begin{equation}
\frac{e^{\ast }H_{c2}^{\perp }}{\hbar c}\left( 
\begin{array}{c}
\frac{3}{2}\left( u_{T}+u_{L}\right) \\ 
-\sqrt{3\left( u_{T}^{2}+u_{L}^{2}\right) -2u_{T}u_{L}}%
\end{array}%
\right) =\alpha \left( T_{c}-T\right) \text{.}  \label{eigenvalue}
\end{equation}%
The corresponding critical field $H_{c2}^{\perp }$ ("perpendicular" refers
to the order parameter direction) that can be expressed via an effective
"perpendicular" coherence length, 
\begin{eqnarray}
H_{c2}^{\perp } &=&\frac{\Phi _{0}}{2\pi \xi _{\perp }^{2}};  \label{Hc2perp}
\\
\xi _{\perp }^{2} &=&\frac{3}{2}\left( \xi _{L}^{2}+\xi _{T}^{2}\right) -%
\sqrt{3\xi _{L}^{4}+3\xi _{T}^{4}-2\xi _{L}^{2}\xi _{T}^{2}}\text{.}
\end{eqnarray}%
It is always larger than $H_{c2}^{\parallel }$, and therefore is physically
realized. The upper field $H_{c2}^{\perp }$ becomes infinite at $%
r_{c}=u_{L}/u_{T}=\left( 13-4\sqrt{10}\right) /3\simeq 0.117$. This means
that in such material superconductivity persists at any magnetic field like
in some $p$- wave superconductors. It was found in Section IV that for the
simplest Dirac semi-metal, $r=1/32<r_{c}$ see Eq.(\ref{u_res}). Thus there
is no upper critical field in this case. Of course, different microscopic
models that belong to the same universality class, might have higher $r$. In
any case the Abrikosov lattice is expected to be markedly different from the
conventional one and even from the vector order parameter model studied in 
\cite{Knigavko}. At large fields in principle there is a direct interaction
between the external magnetic field and spin, not taken into account in
Hamiltonian Eq.(\ref{kinetic}) studied so far. The next subsection addresses
this question.

\subsection{What impact has the Pauli paramagnetism at strong fields?}

Since the prediction of the FFLO effect\cite{Ketterson} in low $T_{c}$
superconductors it is well known that at very high magnetic fields the
direct spin - magnetic field coupling on the microscopic level might not be
negligible. The singlet channel Cooper pair is effectively "broken" by the
splitting since the spins of the two electrons are opposite (Pauli
paramagnetic limit). It is not clear what impact it has on Dirac
semi-metals. If the impact is large it could be incorporated as an
additional paramagnetic term in the GL energy. In an isotropic Dirac
superconductor one has only one possible term in the GL energy term linear
in paramagnetic coupling and consistent with symmetries:%
\begin{equation}
F_{par}=N\mu _{p}\int_{\mathbf{r}}i\left( \mathbf{\Delta }^{\ast }\mathbf{%
\times \Delta }\right) \cdot \mathbf{B}\text{,}  \label{FZ}
\end{equation}%
where $\mu _{p}$ is the effective "spin" of the Cooper pair sometimes called
"Zeeman coupling"\cite{Knigavko,Alama}.

The single particle Hamiltonian in magnetic field with the Pauli term
becomes 
\begin{equation}
\text{\ }\widehat{K}=-iv_{F}\hbar \mathcal{D}\mathbf{\cdot \alpha }-\mu +\mu
_{B}\mathbf{\Sigma }\cdot \mathbf{B}\text{,}  \label{KZeeman}
\end{equation}%
where the Bohr magneton, $\mu _{B}=e\hbar /2mc$, determines the strength of
the coupling of the spin to magnetic field, with $m$ being the free electron
mass. The direct calculation, see Appendix A, shows that $\mu _{p}=0$.

\section{ Discussion and conclusions}

Starting from the microscopic model of the isotropic Dirac semi-metal, the
Ginzburg-Landau energy for this field is derived using the Gor'kov
technique. It was found that the transversal coherence length $\xi _{T}$ is
much smaller than the longitudinal, $\xi _{L}=4\sqrt{2}\xi _{T},$ despite
the isotropy. Several new features appear when an external field is applied.
The Ginzburg - Landau model is used to investigate magnetic properties of
such superconductors. Using the small deviation method the magnetic
penetration depth was found also to be vastly different for longitudinal and
transverse fluctuations $\lambda _{T}/\lambda _{L}=4\sqrt{2}$.

We have observed that properties of the triplet superconductor phase of the
Dirac semi-metal has extremely unusual features that we would like to
associate qualitatively with the characteristics of the Cooper pair. The
superconducting state generally is a Bose - Einstein condensate of composite
bosons - Cooper pairs, classically described by the Ginzburg - Landau energy
as a functional of the order parameter. In the present case the Cooper boson
is described by a \textit{vector field }$\Delta _{i}\left( \mathbf{r}\right) 
$. In this respect it is reminiscent to phonon and vector mesons in particle
physics\cite{Weinberg}. Vector fields generally have both the orbital and
internal degrees of freedom often called polarization. The internal degree
of freedom might be connected to the "valley" degree of freedom of
constituents of the composite boson. We have provided evidence that the
Cooper pair in DSM has finite orbital momentum, albeit, as will be shown
shortly, the spin magnetic moment is zero. Microscopically the unusual
nature is related to the presence of the valley degeneracy in Dirac
semi-metal. While in a single band superconductor the Pauli principle
requires a triplet Cooper pair to have both odd angular momentum and spin,
it is no longer the case in the Dirac semi-metal.

A massive bosonic vector field in isotropic situation (the case considered
here) generally have distinct transversal and longitudinal polarizations
(massless fields like photons in dielectric do not possess the longitudinal
degree of freedom). The results for coherence lengths $\xi _{T,L}$ and the
penetration depths $\lambda _{T,L}$ in triplet superconductor in DSM
demonstrate pronounced disparity between properties of transverse and
longitudinal polarizations. In conventional "scalar" order parameter
superconductor the Abrikosov ratio\cite{Ketterson}, $\kappa \equiv \lambda
/\xi $, distinguishes between type I ($\kappa <\kappa _{c}=1/\sqrt{2}$) and
type II, where, for example, vortices appear under a magnetic field. In the
present case this separation is ambiguous. There are two quite different
ratios: transversal and the longitudinal, (in respect to director of the
order parameter{\huge \ }$n=\Delta /\left\vert \Delta \right\vert )$.see Eq.(%
\ref{u_res}). 
\begin{eqnarray}
\kappa _{T} &=&\frac{\lambda _{T}}{\xi _{T}}=  \label{kappa} \\
\frac{\hbar c}{e^{\ast }u_{T}}\sqrt{\frac{\beta _{1}+\beta _{2}}{4\pi N}} &=&%
\frac{1}{16}\sqrt{\frac{15\pi ^{3}}{7\zeta \left( 3\right) }}\frac{%
cT_{c}\hbar ^{1/2}}{e^{\ast }\mu v_{F}^{1/2}}; \\
\kappa _{L} &=&\frac{\lambda _{L}}{\xi _{L}}=\text{ }\frac{u_{T}}{u_{L}}%
\kappa _{T}\text{.}  \notag
\end{eqnarray}

For typical DSM one estimates the Fermi velocity and chemical potential \cite%
{Potemski} $v_{F}=c/200$, $\mu =0.2eV$, and with the expected critical
temperature \cite{pressureBiSe}, \cite{DasSarma14} $T_{c}=5K$, one obtains
the transversal Abrikosov ratio $\kappa _{T}=0.08,$ smaller than critical,
while the longitudinal, $\kappa _{L}=2.7$ is larger. This means that
longitudinal fluctuations the material behaves as type II, while response to
the transverse ones is that of a type I superconductor. This has an impact
on transport, optical and magnetic properties of these superconductors.

The vortex physics of strongly type II triplet superconductors of this type
is very rich and some of it has already been investigated in connection with
heavy fermion and other superconductors suspected to possess $p$-wave
pairing. In particular, their magnetic vortices appear as either vector
vortices or so-called skyrmions\cite{Knigavko} - coreless topologically
nontrivial textures. The magnetic properties like the magnetization are very
peculiar and even without a magnetic field the system forms a "spontaneous
flux state". The material therefore can be called a "ferromagnetic
superconductor". The superconducting state develops weak ferromagnetism and
a system of alternating magnetic domains\cite{Bel}. It was noted\cite%
{Knigavko} that the phase is reminiscent to the phase B of superfluid $%
He_{3} $. \cite{Ketterson} (with an obvious distinction that the order
parameter in the later case is neutral rather than charged and tensorial
rather than vectorial). As was shown in section V, the Dirac semi-metal
triplet \ superconductor phase is different in several respects. It is more
like the phase B of superfluid $He_{3}$.

Experimentally the major consequence of the present theoretical
investigation, namely the polarization effect of the vector order parameter
should be pronounced in the AC response of these materials. Recently the AC
response of the disordered superconductor was utilized to probe Goldstone
modes\cite{Cea}. We have demonstrated that they are abundant in the triplet
DSM superconductor.

\section{Appendix A.}

\subsection{Critical temperature calculation}

Starting from equation Eq.(\ref{Tc}) the angle integrations result in (for $%
\mu >>T_{D},T_{c})$

\begin{eqnarray}
\frac{1}{g} &=&T\sum\limits_{np}\frac{\mu ^{2}+\omega _{n}^{2}}{\omega
_{n}^{4}+\left( p^{2}-\mu ^{2}\right) ^{2}+2\omega _{n}^{2}\left( p^{2}+\mu
^{2}\right) }  \label{A1} \\
&=&\frac{\mu ^{2}}{12\pi ^{2}}\int_{\varepsilon =-T_{D}}^{T_{D}}\frac{\tanh
\left( \varepsilon /2T\right) }{\varepsilon }\approx \frac{\mu ^{2}}{6\pi
^{2}}\log \frac{2T_{D}\gamma _{E}}{\pi T_{c}}.  \notag
\end{eqnarray}%
where $\varepsilon =v_{F}p-\mu $. See the last (BCS) integral in\cite{AGD}.

\subsection{ Cubic terms coefficients calculation}

To fix the two coefficients, $\beta _{1}$ and $\beta _{2}$ in Eq.(\ref{GLEZ}%
) we use only two components. The particular case $j=k=l=1$ (the coefficient
of $\psi _{1}^{\ast 2}\psi _{1}$) gives after angle integration

\begin{eqnarray}
\beta _{1}+\beta _{2} &=&\frac{2T}{15\pi ^{2}}\sum\limits_{n}\int_{p=0}^{%
\infty }  \label{A2} \\
&&\frac{p^{2}\left( p^{4}+10p^{2}\left( \omega _{n}^{2}-5\mu ^{2}\right)
-15\left( \mu ^{2}+\omega _{n}^{2}\right) ^{2}\right) }{\left(
p^{4}+2p^{2}\left( \omega _{n}^{2}-\mu ^{2}\right) +\left( \mu ^{2}+\omega
_{n}^{2}\right) ^{2}\right) ^{2}}\text{.}
\end{eqnarray}%
Performing finite integration (the upper bound on momentum, $\mu +T_{D}$,
can be replaced by infinity), one obtains 
\begin{equation}
\beta _{1}+\beta _{2}=\frac{8\mu ^{2}}{15\pi ^{4}}s_{3}\text{,}  \label{A3}
\end{equation}%
where the sum is

\begin{equation}
s_{3}=\sum_{n=0}\frac{1}{\left( 2n+1\right) ^{3}}=\frac{7\zeta \left(
3\right) }{4}\text{.}  \label{A4}
\end{equation}%
Similarly taking $j=l=2,k=1$ (the coefficient of $\psi _{2}^{\ast 2}\psi
_{1} $) gives after the angle integration 
\begin{eqnarray}
\beta _{2} &=&\frac{2T}{15\pi ^{2}}\sum\limits_{n}\int_{p=0}^{\infty }
\label{A5} \\
&&\frac{p^{2}\left( 7p^{4}+10p^{2}\left( \omega _{n}^{2}+3\mu ^{2}\right)
+15\left( \mu ^{2}+\omega _{n}^{2}\right) ^{2}\right) }{\left(
p^{4}+2p^{2}\left( \omega _{n}^{2}-\mu ^{2}\right) +\left( \mu ^{2}+\omega
_{n}^{2}\right) ^{2}\right) ^{2}} \\
&=&-\frac{4\mu ^{2}}{15\pi ^{4}T^{2}}s_{3}\text{.}
\end{eqnarray}%
resulting in Eq.(\ref{betas}).

\subsection{Effect of the Pauli interaction}

The single particle Hamiltonian in magnetic field was written in Eq.(\ref%
{KZeeman}). In order to fix the coefficient of the paramagnetic term linear
in both the order parameter and Pauli coupling it is enough to expand the
linearized Gorkov equations Eq.(\ref{linear1}) to the first order in the
spin density. Normal Greens functions have the following corrections: 
\begin{eqnarray}
D_{Z} &\approx &D-\mu _{B}D\left( \Sigma \cdot B\right) D\text{;}  \label{A6}
\\
\widetilde{D}_{Z} &\approx &\widetilde{D}+\mu _{B}\widetilde{D}\left( \Sigma
^{t}\cdot B\right) \widetilde{D}\text{.}  \notag
\end{eqnarray}%
The Pauli term in Gor'kov equation (after multiplying by $T_{i}^{t}$ and
taking the trace as usual), Eq.(\ref{GLEZ}), therefore is obtained from
expansion of Eq.(\ref{linear1}), 
\begin{eqnarray}
&&\sum_{\omega \mathbf{p}}\text{Tr}\left\{ T_{i}\widetilde{D}\Delta ^{\ast
}D\right\}  \label{A7} \\
&=&-i\mu _{Z}\varepsilon _{ijk}\Delta _{j}^{\ast }B_{k}N\left( \mu \right)
=\mu _{B}\mathcal{B}_{ijk}^{Z}\Delta _{j}^{\ast }B_{k}, \\
\mathcal{B}_{ijk}^{Z} &=&\sum_{\omega \mathbf{p}}\text{Tr}\left\{ T_{i}^{{}}%
\widetilde{D}\left( \Sigma _{k}^{t}\widetilde{D}T_{j}^{\ast }-T_{j}^{\ast
}D\Sigma _{k}\right) D\right\} \text{,}  \notag
\end{eqnarray}%
The bubble sum is directly evaluated and vanishes $B_{ijk}^{Z}=0$.

\section{Appendix B. Calculation of gradient terms}

Rotational invariance allows to represent the sum in Eq (\ref{nonlocal _
terms}) terms of coefficients $u_{T}$ and $u_{L}$:%
\begin{eqnarray}
&&-N\left( \mu \right) \left( u_{T}\left( P^{2}\delta
_{mj}-P_{m}P_{j}\right) +u_{L}P_{m}P_{j}\right)  \label{B1} \\
&=&P_{k}P_{l}\sum_{\omega \mathbf{q}}Tr\left\{ T_{m}^{t}\widetilde{D}\left(
\omega ,\mathbf{q}\right) T_{j}^{t\ast }D_{kl}^{\prime \prime }\left( \omega
,\mathbf{q}\right) \right\} ,
\end{eqnarray}%
where 
\begin{eqnarray}
D_{ij}^{\prime \prime } &=&\frac{2}{\left( q^{2}-\left( i\omega +\mu \right)
^{2}\right) ^{2}}  \label{B2} \\
&&\left\{ q_{j}\alpha _{i}+q_{i}\alpha _{j}+\delta _{ij}\left( i\omega +\mu +%
\mathbf{\alpha \cdot q}\right) +2q_{i}q_{j}D\right\} \text{.}
\end{eqnarray}%
In particular

\begin{eqnarray*}
N\left( \mu \right) u_{L} &=&-\sum_{\omega \mathbf{p}}Tr\left\{ T_{z}%
\widetilde{D}\left( \omega ,\mathbf{p}\right) T_{z}^{\ast }D_{zz}^{\prime
\prime }\left( \omega ,\mathbf{p}\right) \right\} \\
&=&\frac{\mu ^{2}}{15\pi ^{4}T^{2}v_{F}\hbar }s_{3} \\
&=&\frac{7\zeta \left( 3\right) }{60\pi ^{4}}\frac{\mu ^{2}}{T^{2}v_{F}\hbar 
}=\frac{7\zeta \left( 3\right) \left( v_{F}\hbar \right) ^{2}N\left( \mu
\right) }{120\pi ^{2}T^{2}}
\end{eqnarray*}%
and%
\begin{equation*}
N\left( \mu \right) u_{T}=-\sum_{\omega \mathbf{p}}Tr\left\{ T_{x}\widetilde{%
D}\left( \omega ,\mathbf{p}\right) T_{x}^{\ast }D_{zz}^{\prime \prime
}\left( \omega ,\mathbf{p}\right) \right\} =32u_{L}\text{.}
\end{equation*}

\section{Appendix C. Exact solution for upper critical magnetic field}

In this Appendix the matrix $\mathcal{H}$ defined in Eq.(\ref{2DVar})
determining the perpendicular upper critical field is diagonalized
variationally.

\subsection{Creation and annihilation operators}

Using Landau creation and annihilation operators in units of magnetic length 
$\frac{e^{\ast }B}{c}=l^{-2}$ for the state with $k_{x}=0$ (independent of $%
x $), so that covariant derivatives are 
\begin{eqnarray}
D_{x} &=&\partial _{x}+iy=iy=\frac{i}{\sqrt{2}}\left( a+a^{+}\right) ;
\label{C1} \\
D_{y} &=&\partial _{y}=\frac{1}{\sqrt{2}}\left( a-a^{+}\right) .  \notag
\end{eqnarray}%
In terms of these operators the matrix operator $\mathcal{H}$ takes a form:%
\begin{eqnarray}
\mathcal{H} &=&u_{T}+\frac{u}{2}+\mathcal{V};  \label{C2} \\
\mathcal{V}_{11} &=&2u_{T}a^{+}a+\frac{u}{2}\left(
a^{2}+a^{+2}+2a^{+}a\right) \\
\mathcal{V}_{12} &=&\mathcal{V}_{21}=\frac{iu}{2}\left( a^{+2}-a^{2}\right)
\\
\mathcal{V}_{22} &=&2u_{T}a^{+}a-\frac{u}{2}\left(
a^{2}+a^{+2}-2a^{+}a\right)
\end{eqnarray}

The exact lowest eigenvalue is a combination of two lowest Landau levels.
Indeed applying the operator $\mathcal{V}$ on a general vector on the
subspace gives 
\begin{eqnarray}
&&\mathcal{V}\left( 
\begin{array}{c}
\alpha \left\vert 0\right\rangle +\beta \left\vert 2\right\rangle \\ 
\gamma \left\vert 0\right\rangle +\delta \left\vert 2\right\rangle%
\end{array}%
\right)  \label{C3} \\
&=&\left( 
\begin{array}{c}
\frac{u}{\sqrt{2}}\left( i\delta -\beta \right) \left\vert 0\right\rangle \\ 
-\left( u\left( \frac{\alpha }{\sqrt{2}}+\frac{i\gamma }{\sqrt{2}}+2\beta
\right) +4u_{T}\beta \right) \left\vert 2\right\rangle \\ 
+\frac{u}{2}\left( -\beta -i\delta \right) \left\vert 4\right\rangle \\ 
+\frac{u}{\sqrt{2}}\left( i\beta +\delta \right) \left\vert 0\right\rangle
\\ 
+\left( u\left( -\frac{i\alpha }{\sqrt{2}}+\frac{\gamma }{\sqrt{2}}-2\delta
\right) -4u_{T}\delta \right) \left\vert 2\right\rangle \\ 
+\frac{u}{2}\left( -i\beta +\delta \right) \left\vert 4\right\rangle%
\end{array}%
\right) \text{.}  \notag
\end{eqnarray}%
For $\delta =i\beta $, higher Landau levels decouple and one gets eigenvalue
equations%
\begin{equation*}
\left\vert 
\begin{array}{ccc}
-v & -\sqrt{2}u & 0 \\ 
-\frac{u}{\sqrt{2}} & -4u_{T}-2u-v & -\frac{iu}{\sqrt{2}} \\ 
0 & ui\sqrt{2} & -v%
\end{array}%
\right\vert =0\text{,}
\end{equation*}%
resulting in three eigenvalues of $\mathcal{H}$ 
\begin{eqnarray}
h^{\left( 1\right) } &=&u_{T}+u/2,h^{\left( \pm \right) }  \label{C5} \\
&=&3u_{T}+\frac{3}{2}u\pm \sqrt{4u_{T}^{2}+4u_{T}u+3u^{2}}\text{.}
\end{eqnarray}

\textit{Acknowledgements.} We are indebted to D. Li and C. W. Luo for
explaining details of experiments, and M. Lewkowicz for valuable
discussions. Work of B.R. was supported by NSC of R.O.C. Grants No.
98-2112-M-009-014-MY3 and MOE ATU program.

\end{document}